\documentstyle[prb,aps,preprint]{revtex}
\begin{document}

\draft
\title{Asymptotic results on the product of random probability 
matrices }

\author{X. R. Wang}
\address{Physics Department,
The Hong Kong University of Science and Technology,}
\address{ Clear Water Bay,
Hong Kong.
}
\maketitle
\begin{abstract}
I study the product of independent identically distributed $D\times D$ 
random probability matrices. Some exact asymptotic results are obtained. 
I find that both the left and the right products approach exponentially 
to a probability matrix(asymptotic matrix) in which any two rows are the 
same. A parameter $\lambda$ is introduced for the exponential coefficient 
which can be used to describe the convergent rate of the products. 
$\lambda$ depends on the distribution of individual random matrices. I 
find $\lambda = 3/2$ for $D=2$ when each element of individual random 
probability matrices is uniformly distributed in $[0,1]$. In this case, 
each element of the asymptotic matrix follows a parabolic distribution 
function. The distribution function of the asymptotic matrix elements can be
numerically shown to be non-universal.
Numerical tests are carried out for a set of random 
probability matrices with a particular distribution function. 
I find that $\lambda $ increases  monotonically from $\simeq 1.5$ to 
$\simeq 3$ as $D$ increases from $3$ to $99$, and the distribution of 
random elements in the asymptotic products can be described by a Gaussian
function with its mean to be ${{1}\over{D}}$.
\end{abstract}
{\bf Key words:} random probability matrix,  product, asymptotic behavior 
\pacs{PACS: 02.10, 02.50, 05.20}

In recent years, there is an increasing interest in studying the 
properties of random matrices\cite{book}.
Random matrices can be used to describe 
disordered systems, chaos, and biological problems, and in the 
statistical description of complex nuclei\cite{book,rma}.
There are two different kind of problems related to random matrices. One of 
them is to study the statistical properties of a single random matrix.
Wigner\cite{epw} was the first  
to use a single large random matrix to explain the statistical behavior of 
levels in nuclear physics. The semicircular and the circular 
theorems\cite{book,epw} had been obtained for the Gaussian unitary, Gaussian 
orthonormal, and Gaussian symplectic ensembles. The second problem is to 
study the statistical behavior of a product of random matrices. The product of 
random matrices has attracted much attention in most of recent works 
because many statistical problems in disordered systems and 
chaotic dynamical systems can be formulated as a study of such product. 
For the product of random matrices, there are beautiful 
Furstenberg\cite{fk} and Oseledec\cite{vio} 
theorems about the existence of Lyapunov 
characteristic exponents. However, a detailed analysis of the product of 
random matrices is very difficult due to the noncommutability of random 
matrices, and there are not so many general exact results about structures 
of the products of random matrices. Therefore, any exact results on the product 
of a particular type of random matrices should be interesting. 

In this work, I present some exact results of the product of independent 
identically distributed random probability matrices. A probability matrix $T$ 
of $D \times D$ is defined as $T(ij) \ge 0$ and $\sum_{j=1}^{D} 
T(ij)=1$, $i=1,2,...D$, where $T(ij)$ are the elements of $T$. 
Physically, such matrices can be used to describe the dynamics of a $D$ 
state classical system if $T_{t}(ij)$ is interpreted as the probability for 
the system in $i'th$ state jumping to $j'th$ state at time $t$. If the 
hopping process is stochastic,  the evolution of the system is described 
by the product of the random probability matrices $T_{t}$. 
It is easy to show that the product of two probability matrices is still a 
probability matrix, ie. $A=T_1T_2$ is a probability matrix provided $T_1$ 
and $T_2$ are two probability matrices of order $D\times D$. In this paper, 
I show that the product of independent identically distributed random 
probability matrices approaches exponentially, in terms of the number 
of matrices in the product, to a matrix in which any two rows are the same. 
For $2\times 2$ independent random probability matrices 
in which any element is uniformly distributed in $[0,1]$, I find that 
the asymptotic matrix elements have parabolic 
distribution. The main results are
described by the following propositions.

{\bf Proposition 1:} Let $\{ T_{k} \}$ be a set of $D \times D$ independent 
identically distributed random probability matrices in which all elements 
are random and have the same distribution function. If $A(n)=\prod_{k=1}^{n} 
T_{k} \equiv T_{n}T_{n-1}\cdot\cdot\cdot T_{2}T_{1}$(the left product) and 
$B(n)=\prod_{k=1}^{n} T_{k} \equiv T_{1}T_{2}\cdot\cdot\cdot T_{n-1}T_{n}
$(the right product), then
\begin{equation}
\label{left}
\lim_{n\to \infty} A(n)=
\left( \begin{array}{cccc}
a(1)&a(2)&\ldots&a(D)\\
a(1)&a(2)&\ldots&a(D)\\
\vdots&\vdots&\ddots&\vdots\\
a(1)&a(2)&\ldots&a(D)
\end{array} \right)
\end{equation}
and 
\begin{equation}
\label{right}
\lim_{n\to \infty} B(n)=
\left( \begin{array}{cccc}
a(1)&a(2)&\ldots&a(D)\\
a(1)&a(2)&\ldots&a(D)\\
\vdots&\vdots&\ddots&\vdots\\
a(1)&a(2)&\ldots&a(D)
\end{array} \right),
\end{equation}
where $a(i)$, $(i=1,2,\ldots D)$, are positive random numbers with $\sum_{i} 
a(i)=1$. However, the values of $a's$ in the left product are fixed for a 
given sequence of random matrices while they keep changing in 
the right product. 

Before we prove this proposition, let us look at a special case of $D=2$. 
Let 
\begin{equation}
\label{2by2}
A(n)\equiv T_{n}\cdot\cdot\cdot T_{2}T_{1}=
\left( \begin{array}{cc}
y_{n}(1)&1-y_{n}(1)\\
y_{n}(2) &1-y_{n}(2)
\end{array} \right),
\end{equation}
then, from $A(n)=T_{n}A(n-1)$, we obtain the following recursion relations 
for $y_{n}(1)$ and $y_{n}(2)$
\begin{mathletters}
\begin{equation}
\label{y1}
y_{n}(1)=x(1)y_{n-1}(1)+(1-x(1))y_{n-1}(2), 
\end{equation}
\begin{equation}
\label{y2}
y_{n}(2)=x(2)y_{n-1}(1)+(1-x(2))y_{n-1}(2), 
\end{equation}
\end{mathletters}
where $x(1)$ and $x(2)$ are the two independent random elements of $T_n$, i.e.,
\[
T_{n}=
\left( \begin{array}{cc}
x(1)&1-x(1)\\
x(2) &1-x(2)
\end{array} \right).
\]
Therefore,
\[
y_{n}(1)-y_{n}(2)=[x(1)-x(2)][y_{n-1}(1)-y_{n-1}(2)]
\]
which gives 
\begin{equation}
\label{dy2}
|{{y_{n}(1)-y_{n}(2)}\over{y_{n-1}(1)-y_{n-1}(2)}}|=|x(1)-x(2)|\le 1. 
\end{equation}
Thus, we expect $|y_{n}(1)-y_{n}(2)|$ approaches to zero exponentially. If 
$x(1)$ and $x(2)$ are uniformly distributed in $[0,1]$, then, for a large $n$,
\begin{equation}
\label{dy3}
y_{n}(1)-y_{n}(2)\sim e^{-\lambda n}, 
\end{equation}
with 
\[
\lambda=-<ln|x(1)-x(2)|>=-\int_{0}^{1}\int_{0}^{1} 
ln|x(1)-x(2)| dx(1) dx(2)=3/2. 
\]
Therefore, the left product 
of independent identically distributed $2\times 2$
random probability matrices exponentially approaches to a matrix of the form
\begin{equation}
\label{dy4}
\left( \begin{array}{cc}
a_1&1-a_1\\ a_2 &1-a_2
\end{array} \right).
\end{equation}
with $\lambda = 1.5$. Similarly, it is easy to show that the same 
conclusion can be made for a right product.

The above approach can be extended to the general cases. Without losing 
generality, we need only show that the values of elements in the first column 
of the product of random probability matrices approach to each other. Let 
\begin{equation}
\label{dbyd1}
A_{n} \equiv 
\left( \begin{array}{ccccc}
y_{n}(1)&\ &\ &\ &\ \\
y_{n}(2)&\ &\ &\ &\ \\ 
\dots &\ &\ast &\ &\ \\ 
\ldots &\ &\ &\ &\ \\ 
y_{n}(D)&\ &\ &\ &\ 
\end{array} \right)= 
T_{n}A(n-1). 
\end{equation}
We want to show that $<|y_{n}(k)-y_{n}(1)|> \simeq \exp^{-\lambda n}$, 
$\lambda > 0. $ It is not hard to see that 
\begin{equation}
\label{dy5}
y_{n}(i)-y_{n}(D)=\sum_{k=1}^{D-1}[x(i,k)-x(D,k)][y_{n-1}(k)-y_{n-1}(D)], 
\end{equation}
where $x(i,j)$ are matrix elements of $T_n$ which satisfies the conditions of 
a probability matrix. $\sum_{j} x(i,k)=1$ is used in the above 
derivation. Define 
\begin{equation}
\label{dy6}
z_{n}(i)=y_{n}(i)-y_{n}(D) \qquad \quad i=1,\ldots,D-1, 
\end{equation}
equation (\ref{dy5}) can be written in the following matrix form
\[
\left( \begin{array}{c}
z_{n}(1)\\ z_{n}(2)\\ \vdots\\ z_{n}(D-1)
\end{array} \right)=C_n
\left( \begin{array}{c}
z_{n-1}(1)\\ z_{n-1}(2)\\ \vdots\\ z_{n-1}(D-1)
\end{array} \right)=
\]
\begin{equation}
\label{dy7}
\left( \begin{array}{ccc}
x(1,1)-x(D,1) &\ldots &x(1,D-1)-x(D,D-1)\\ 
x(2,1)-x(D,1) &\ldots &x(2,D-1)-x(D,D-1)\\
\vdots &\ddots &\vdots \\ 
x(D-1,1)-x(D,1) &\ldots &x(D-1,D-1)-x(D,D-1)
\end{array} \right)
\left( \begin{array}{c}
z_{n-1}(1)\\ z_{n-1}(2)\\ \vdots\\ z_{n-1}(D-1)
\end{array} \right)
\end{equation}
where $C_n$ is a $(D-1)\times (D-1)$ matrix related to probability matrix $T_n$.
It is not difficulty to show the following relation between $T_{n}$ and $C_n$ 
\begin{equation}
\label{c1}
\Vert T_{n}-\mu {\bf I} \Vert=(1-\mu ) \Vert C_n - \mu {\bf I}\Vert,
\end{equation}
that is the eigenvalues of matrix $C_n$ are $D-1$ of eigenvalues of 
matrix $T_n$ (one of the eigenvalues of $T_n$ with 
value $1$ is excluded)\cite{note}. It is well known that the magnitudes of 
eigenvalues $\mu_{i},\ i=1,\ldots, D$, of a $D\times D$ probability matrix 
are not greater than $1$, i.e. $| \mu_{i} | \le 1$. Furthermore, matrices
$C's$ do not have any common eigenvectors\cite{note2}. Therefore,
\begin{equation}
\label{c2}
<\left( \begin{array}{c}
z_{n}(1)\\ z_{n}(2)\\ \vdots\\ z_{n}(1)
\end{array} \right)>
\sim e^{-\lambda N}
\qquad (for \  a \ large \ N) 
\end{equation}
with $\lambda > 0$, i.e., the values of elements in the first 
column of the left product $A$ approach exponentially to each other. It 
is easy to show that the same result is true for any other column of $A$. 
Thus, relation (\ref{left}) holds. Similarly, it can be shown that relation 
(\ref{right}) also holds. 

This result is not really surprising. It is known that the magnitudes of 
eigenvalues of a probability matrix are equal to or smaller than 
$1$. A probability matrix always has an eigenvalue 1 with the corresponding 
eigenvector(mode) $\pmatrix{1\cr 1\cr \vdots\cr 1\cr}$ (There may 
exist other eigenvectors with eigenvalue $1$). The eigenvalues of the 
product will either approach to $0$ or stay at $1$ when such matrices 
are multiplied together. Because each matrix is random and independent, 
these matrices are not commutable among themselves, and they do not 
in general have the same eigenvector except 
$\pmatrix{1\cr 1\cr \vdots\cr 1\cr}$ which will keep 
unchanged since its eigenvalue is equal to $1$. Therefore, all other 
modes are mixed together, and decay with the multiplication. The relation 
(\ref{left}) is then expected. 

{\bf Proposition 2:} Let $\{ T_{k} \}$ be a set of $2 \times 2$ independent 
identically distributed random probability matrices in which all elements 
are uniformly distributed in $[0,1]$, proposition 1 guarantees that 
\[
\lim_{n\rightarrow \infty} A(n)\equiv \lim_{n\rightarrow \infty} 
T_{n}T_{n-1}\cdot\cdot\cdot T_{2}T_{1}= 
\left( \begin{array}{cc}
a & 1-a\\
a & 1-a
\end{array} \right)
\]
and
\[
\lim_{n\rightarrow \infty } B(n)\equiv \lim_{n\rightarrow \infty}
T_{1}T_{2}\cdot\cdot\cdot T_{n-1}T_{n}= 
\left( \begin{array}{cc} 
a & 1-a\\
a & 1-a
\end{array} \right) 
\]
Then $a$ is a random variable whose distribution function is 
\begin{equation}
\label{fa}
f(a)=6a(1-a).
\end{equation}

To prove proposition 2, we notice that $a$ is a 
random variable which obeies recursion relation
\begin{equation} 
\label{recursion} 
a_{n}=x(1)a_{n-1}+x(2)(1-a_{n-1}),
\end{equation}
where $x(1)$ and $x(2)$ are two indenpendent random numbers uniformly 
distributed in $[0,1]$. Since $f(a)$ is the asymptotic distribution 
function of $a_n$, i.e., $n\rightarrow \infty$, $a_n$ and $a_{n-1}$ 
should have the same distribution function $f(a)$ when $n\rightarrow \infty$.
Therefore, we can 
obtain the following equation in the integral form for distribution 
function $f(a)$ 
\begin{equation}
\label{fa1}
f(a)=\int_{0}^{1} \int_{0}^{1} \int_{0}^{1} f(b) 
\delta(a-bx_1+bx_2-x_2) d x_1 dx_2 db.
\end{equation}
Substitute $\delta(a-bx_1+bx_2-x_2)$ by 
\begin{equation}
\label{delta}
\frac{1}{2\pi} \int_{-\infty}^{\infty} e^{iq(a-bx_1+bx_2-x_2)}dq 
\end{equation}
and integrate over $x_1$ and $x_2$, equation (\ref{fa1}) becomes 
\begin{equation}
\label{fa2}
f(a)=\frac{1}{2\pi}\int_{0}^{1}db \frac{f(b)}{b(b-1)} \int_{-\infty}^{\infty}
dq\frac {e^{iqa}+e^{iq(a-1)}-e^{iq(a-b)}-e^{iq(a+b-1)}}{q^2}.
\end{equation}
Differentiate equation (\ref{fa2}) with respect to $a$ twice and notice 
$f(a)=f(1-a)$, we can show 
that $f(a)$ satisfies differential equation 
\begin{equation}
\label{fa3}
f''=-\frac {2} {a(1-a)}f, 
\end{equation}
with boundary conditions $f(0)=f(1)=0$.
Equation (\ref{fa3}) can be easily solved by power-series expansion method 
since $a=0$ (or $a=1$) is a regular singular point.  
The solution of this equation (normalize to 1) is that of equation 
(\ref{fa}).

I have carried out some numerical simulations to further confirm the results 
in the above propositions. Figure \ref{distribution1} is the 
distribution function of an element of the asymptotic matrix of the 
product of $2\times 2$ independent random probability matrices in 
which elements are uniformly distributed in $[0,1]$. Solid line is the 
numerical result, and dashed line is the analytical expression (\ref{fa}). 
They agree very well with each other. In order to check whether the 
distribution function of the product depends on the distribution 
function of individual random probability matrices, I also study the 
product of independent random probability matrices in 
which elements are distributed in $[0,1]$ according to function
\begin{equation}
\label{fa4}
f(x)=\int_{0}^{1} \ldots \int_{0}^{1} \delta(x- \frac {x_1}
{\sum_{1}^{D} x_{i}}) dx_{1}\ldots dx_{D}.
\end{equation}
Distribution (\ref{fa4}) is chosen because it is easy to generate 
on a computer. Although I cannot find the distribution function for the 
matrix elements of the asymptotic product analytically in this case, 
numerical results can easily be obtained. Figure \ref{distribution2} 
is the single variable distribution functions of random matrix elements 
in the right product of such random probability matrices of order of 
$2\times 2$ and $4\times 4$. The dashed lines are the numerical results,  
and the solid lines are the fits of Gaussian functions. The numerical 
results can be well described by a Gaussian function with its mean 
to be ${{1}\over{D}}$.
Compare with that in figure \ref{distribution1}, we can see that the 
distribution function of the product of independent identically distributed
random probability matrices depends on the distribution of individual 
random matrices. In other words, unlike the large number theorem for random
numbers, the distribution function of the product is not universal. 

I check numerically the results in proposition 1 by using the random 
probability matrices whose elements are independently, except the constrains
of a probability matrix, distributed in $[0,1]$ according to equation 
(\ref{fa4}). I compute the decay of $<|A_{n}(1,1)-A_{n}(2,1)|>$ with $n$, 
where $<\ldots>$ denotes ensemble average, and $A_{n}(i,j)$ 
are the elements of the product of $n$ random matrices. 
Figure \ref{decayl} is $\ln <|A_{n}(1,1)-A_{n}(2,1)|>$ vs. $\ln n$ for 
the left product of $3\times 3$, $4\times 4$, $8\times 8$, 
$16\times 16$, and $99\times 99 $ random matrices. 
100 ensembles are used in the numerical study. 
Figure \ref{decayr} is a similar plot (as figure 1) for a 
right product. The exponential decay of the quantity is clearly shown in 
these figures. Numerically, I find that $\lambda $ increases 
monotonically from $\simeq 1.5$ to $\simeq 3$ as $D$ increases 
from $3$ to $99$. 

In conclusion, I have shown that both left and right products of a sequence of
independent identically distributed random probability matrices exponentially 
approach to a probability matrix in which all elements in any column 
vector are the same. An exponential exponent is used to describe this 
approach rate. I also find that $\lambda $ increases  monotonically 
from $\simeq 1.5$ to $\simeq 3.$ as $D$ increases from $3$ to $99$ 
when the distribution function of individual random matrices is 
described by equation (\ref{fa4}). I also find that $\lambda = 3/2$ for 
$D=2$ when random matrix elements are uniformly distributed in $[0,1]$. 
It is well known that at least one of the eigenvalues of a probability 
matrix is equal to $1$ while the rest of them are distributed in 
$[0,1]$\cite{fw}. A large $D$  means that the product has 
more channel to decay to the stable structure (\ref{left}) 
or (\ref{right}). Thus it is expected that the decay is 
faster for large $D$, ie., $\lambda$ increases 
with $D$. In order to understand the meaning of the results, let us 
look at a physical model system of $D$ states. Assume the system can move  
randomly from $i'$th state to $j'$th state with probability $T_{t}(ij)$ 
at time step $t$. If the system starts from an initial distribution, one 
might want to know the probability of the system in $i'$ state, i.e., the 
distribution function, after a long time. The questions may be whether 
there is a stable distribution (equilibrium state), and/or what it is if 
there is one. Obviously, the long time distribution(s) are the non-trivial 
left eigenstate(s) of the right product of the random matrices $T_{t}$. 
In equation (\ref{right}), although the structure for a product will not change 
when $n$ is larger than certain value, elements $a(i)$ do change as 
another independent random probability matrix is multiplyed to the product. 
Therefore, the system does not have a stable distribution as one expected 
since the dynamics of the system is a stochastic process, and transition 
probabilities keep change with time. Both of the right and left products, 
however, have a non-trivial unique right eigenstate with eigenvalue $1$. The 
eigenstate takes the same value in each of its $D$ components. Unfortunately, 
I am not able to obtain any meaningful non-trivial results by applying the 
propositions to this simple model system. It will be interesting to find 
some interesting physical systems in which the propositions can be used to 
extract useful information. In contrast to the the sum of independent 
random variables whose distribution is Gaussian no matter what is the 
distribution function of individual random variables, the distribution 
function of a element in the product of independent identically 
distributed random matrices depends on the distribution of individual 
random matrices. Therefore, the distribution function of the product of 
independent identically distributed random matrices is not universal.  

The author would like to point out that proposition 2 results from the 
fruitful discussions with Prof. B. Derrida. Prof. Derrida made the major 
contribution to the result.  The author thanks Prof. Derrida for useful 
suggestions and for reading the manuscript. The stimulating discussions
with Prof. Y. Shapir, Dr. W. K. Ge, and Prof. E. Domany  are also 
acknowledged. This work was supported 
by UPGC, Hong Kong, through RGC Grant. 

\hfil\vfill\eject

\begin{figure}
\caption{\label{distribution1}
Distribution function of a random matrix elements in the right product
of independent identically distributed $2\times 2$ random probability
matrices in which all elements are uniformly distributed
in $[0,1]$. The dash line is the numerical result, and the solid line is
$f(a)=6a(1-a)$.}
\end{figure}
 
\begin{figure}
\caption{\label{distribution2}
Distribution function of an arbitrary random matrix element in the right
product of independent identically distributed $2\times 2$, $4\times 4$
random probability matrices in which all elements are
distributed in $[0,1]$ according to equation (\protect\ref{fa4}).
The dashed lines are the numerical results, and the solid lines
are the fits of Gaussian functions.}
\end{figure}
 
\begin{figure}
\caption{\label{decayl}
$\ln <|A_{n}(1,1)-A_{n}(2,1)|>$ vs $n$ of the {\bf left}
product of independent identically distributed random matrices of $3\times
3$, $4\times 4$, $8\times 8$, $16\times 16$, $99\times 99$, with 100
$ensembles. \lambda$'s can be obtained from the slopes. The slopes
increase monotonically from $\lambda \simeq 1.5$ to $\lambda \simeq 3.$
as $D$ changes from $D=3$ to $D=99$.}
\end{figure}
 
\begin{figure}
\caption{\label{decayr}
$\ln <|A_{n}(1,1)-A_{n}(2,1)|>$ vs $n$ of the {\bf right}
product of independent identically distributed random matrices of $3\times
3$, $4\times 4$, $8\times 8$, $16\times 16$, $99\times 99$.
$\lambda$'s can be obtained from the slopes. The slopes
increase monotonically from $\lambda \simeq 1.5$ to $\lambda \simeq 3.$
as $D$ changes from $D=3$ to $D=99$.}
\end{figure}
 

\begin{references}
\bibitem{note2} A probability matrix $T_{i}$ always has an eigenstate 
$\pmatrix{1\cr 1\cr \vdots\cr 1\cr}$ with eigenvalue 1. The rest of 
eigenvalues and eigenstates are given by matrix $C_n$ which depends on 
the detail structure of $T_i$. In general, $C_n$ do not have common 
eigenstates.
\end{references}
\end{document}